\documentclass[a4paper]{jpconf}
\usepackage{graphicx}
\begin{document}
\title{Dileptons from the nonequilibrium Quark-Gluon Plasma}

\author{E. L. Bratkovskaya$^a$, O. Linnyk$^a$
and W. Cassing$^b$}

\address{$^a$Institut f\"ur Theoretische Physik, Universit\"{a}t Frankfurt, 60438 Frankfurt am Main,
  Germany \\
  $^b$Institut f\"ur Theoretische Physik, Universit\"at Giessen,
  35392 Giessen, Germany}


\begin{abstract}
According to the dynamical quasiparticle model (DQPM) -- matched to
reproduce lattice QCD results in thermodynamic limit, -- the
constituents of the strongly interacting quark-gluon plasma (sQGP)
are massive and off-shell quasi-particles (quarks and gluons) with
broad spectral functions. In order to address the electromagnetic
radiation of the sQGP, we derive off-shell cross sections of $q\bar
q\to\gamma^*$, $q\bar q\to\gamma^*+g$ and $qg\to\gamma^*q$($\bar q
g\to\gamma^* \bar q$) reactions taking into account the effective
propagators for quarks and gluons from the DQPM. Dilepton production
in In+In collisions at 158~AGeV is studied by implementing these
processes into the parton-hadron-string dynamics (PHSD) transport
approach. The microscopic PHSD transport approach describes the full
evolution of the heavy-ion collision: from the dynamics of
quasi-particles in the sQGP phase (when the local energy density is
above $\sim 1$~GeV/fm$^3$) through hadronization and to the
following hadron interactions and off-shell propagation after the
hadronization. A comparison to the data of the NA60 Collaboration
shows that the low mass dilepton spectra are well described by
including a collisional broadening of vector mesons, while the
spectra in the intermediate mass range are dominated by off-shell
quark-antiquark annihilation, quark Bremsstrahlung and gluon-Compton
scattering in the nonperturbative QGP. In particular, the observed
softening of the $m_T$ spectra at intermediate masses (1~GeV~$\le M
\le$~3~GeV) is approximately reproduced. \end{abstract}

\section{Introduction}

While the  properties of hadrons are rather well known in free space
(embedded in a nonperturbative QCD vacuum) the masses and lifetimes
of hadrons in a baryonic and/or mesonic environment are subject of
current research in order to achieve a better understanding of the
strong interaction and the nature of confinement. A  modification of
vector mesons has been seen experimentally in the enhanced
production of lepton pairs above known sources in nucleus-nucleus
collisions at Super-Proton-Synchroton (SPS) energies
\cite{CERES,HELIOS}. This can be attributed to a shortening of the
lifetime of the vector mesons $\rho$, $\omega$ and $\phi$. The
question arises if the enhancement might (in part) be due to new
radiative channels~\cite{Linnyk:2009nx} from the strong Quark-Gluon
Plasma (sQGP).  The answer to this question is nontrivial due to the
nonequilibrium nature of the heavy-ion reactions and covariant
transport models have to be incorporated to disentangle the various
sources that contribute to the final dilepton spectra seen
experimentally.

\section{The PHSD approach}

To address the vector meson properties in a hot and dense medium --
as created in heavy-ion collisions -- we employ an up-to-date
relativistic transport model, i.e. the Parton Hadron String
Dynamics model \cite{CasBrat} (PHSD) that incorporates the relevant
off-shell dynamics of the vector mesons as well as the explicit
partonic phase in the early hot and dense reaction region. PHSD
consistently describes the full evolution of a relativistic
heavy-ion collision from the initial hard scatterings and string
formation through the dynamical deconfinement phase transition to
the quark-gluon plasma (QGP) as well as hadronization and to the
subsequent interactions in the hadronic phase.

In the hadronic sector PHSD is equivalent to the
Hadron-String-Dynamics (HSD) transport approach
\cite{CBRep98,Brat97,Ehehalt} that has been used for the description
of $pA$ and $AA$ collisions from SIS to RHIC energies and has lead
to a fair reproduction of hadron abundances, rapidity distributions
and transverse momentum spectra. In particular, HSD incorporates
off-shell dynamics for vector mesons -- according to
Refs.~\cite{Cass_off1} -- and a set of vector-meson spectral
functions~\cite{Brat08} that covers possible scenarios for their
in-medium modification. The transport theoretical description of
quarks and gluons in PHSD is based on a dynamical quasiparticle
model for partons matched to reproduce lattice QCD results in
thermodynamic equilibrium (DQPM). The transition from partonic to
hadronic degrees of freedom is described by covariant transition
rates for the fusion of quark-antiquark pairs to mesonic resonances
or three quarks (antiquarks) to baryonic states.

Various models predict that hadrons change in the (hot and dense)
nuclear medium; in particular, a broadening of the spectral function
or a mass shift of the vector mesons has been expected. Furthermore,
QCD sum rules indicated that a mass shift may lead to a broadening
and vice versa~\cite{MuellerSumRules}; therefore  both modifications
should be studied simultaneously. Thus we explore three possible
scenarios:
 (1) a broadening of
the $\rho$ spectral function, (2) a mass shift, and (3) a broadening
plus a mass shift. The HSD (PHSD) off-shell transport approach
allows to investigate in a consistent way the different scenarios
for the modification of vector mesons in a hot and dense medium. In
off-shell transport the hadron spectral functions change
dynamically during the propagation through the medium and evolve
towards the on-shell spectral function in the vacuum.

As demonstrated in Ref.~\cite{Brat08} the off-shell dynamics is
particularly important for resonances with a rather long lifetime in
vacuum but strongly decreasing lifetime in the nuclear medium
(especially $\omega$ and $\phi$ mesons) but also proves vital for
the correct description of dilepton decays  of $\rho$ mesons with
masses close to the two pion decay threshold. For a detailed
description of the various hadronic channels included for dilepton
production as well as the off-shell dynamics we refer the reader to
Refs.~\cite{Brat08,here}.

\begin{figure}
\includegraphics[width=0.9\textwidth]{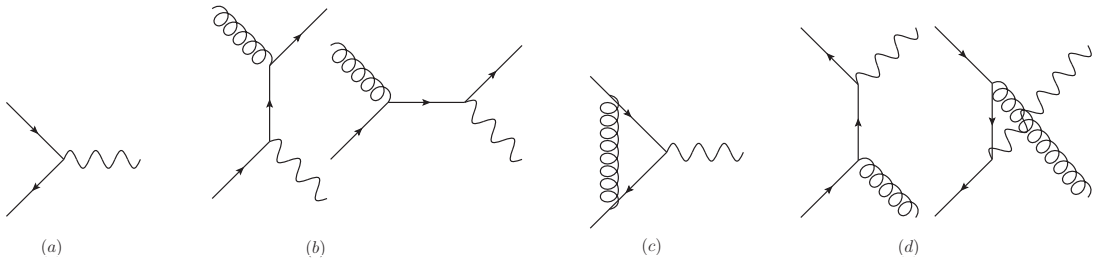}
\caption{Diagrams contributing to the dilepton production from a
QGP: (a) Drell-Yan mechanism, (b) gluon-Compton scattering (GCS),
(c) vertex correction, (d) gluon Bremsstrahlung (NLODY), where
virtual photons (wavy lines) split into lepton pairs, spiral lines
denote gluons, arrows denote quarks. In each diagram the time runs
from left to right.} \label{diagrams}
\end{figure}

\section{SPS energies}

By employing the HSD approach to the low mass dilepton production in
relativistic heavy-ion collisions, it was shown in~\cite{here} that
the NA60 Collaboration data for the invariant mass spectra for
$\mu^+\mu^-$ pairs from In+In collisions at 158 A$\cdot$GeV favored
the 'melting $\rho$' scenario \cite{NA60}. Also the data from the
CERES Collaboration \cite{CERES2} showed a preference for the
'melting $\rho$' picture. On the other hand, the dilepton spectrum
from In+In collisions at 158 A$\cdot$GeV for $M>1$~GeV could not be
accounted for by the known hadronic sources (see Fig.2
of~\cite{here}). Also, hadronic models do not reproduce the
softening of the $m_T$ distribution of dileptons for
$M>1$~GeV~\cite{NA60}. This observation pointed towards a partonic
origin.

Dilepton radiation by the constituents of the strongly interacting
QGP proceeds via the elementary processes illustrated in
Fig.~\ref{diagrams}: the basic Drell-Yan $q+\bar q$ annihilation
mechanism, Gluon Compton scattering ($q+g\to \gamma^*+q$ and $\bar
q+g\to \gamma^*+\bar q$), and quark + anti-quark annihilation with
gluon Bremsstrahlung in the final state ($q+\bar q\to g+\gamma^*$).
In the on-shell approximation, one uses perturbative QCD cross
sections for the processes listed above. The obtained off-shell
elementary cross sections have been implemented into the PHSD
transport code. Note that in our calculations the running coupling
$\alpha_S$ depends on the local energy density $\epsilon$ according
to the DPQM~\cite{Cassing06}, while the energy density is related to
the temperature $T$ by the lQCD equation of state. Numerically, we
observe that $\alpha_S$ is of the order $O(1)$ and thus the
contribution of the higher order diagrams in Fig.~\ref{diagrams} is
not subleading!

\begin{figure*}
  \begin{minipage}[b]{0.54\textwidth}
    \includegraphics[width=\textwidth]{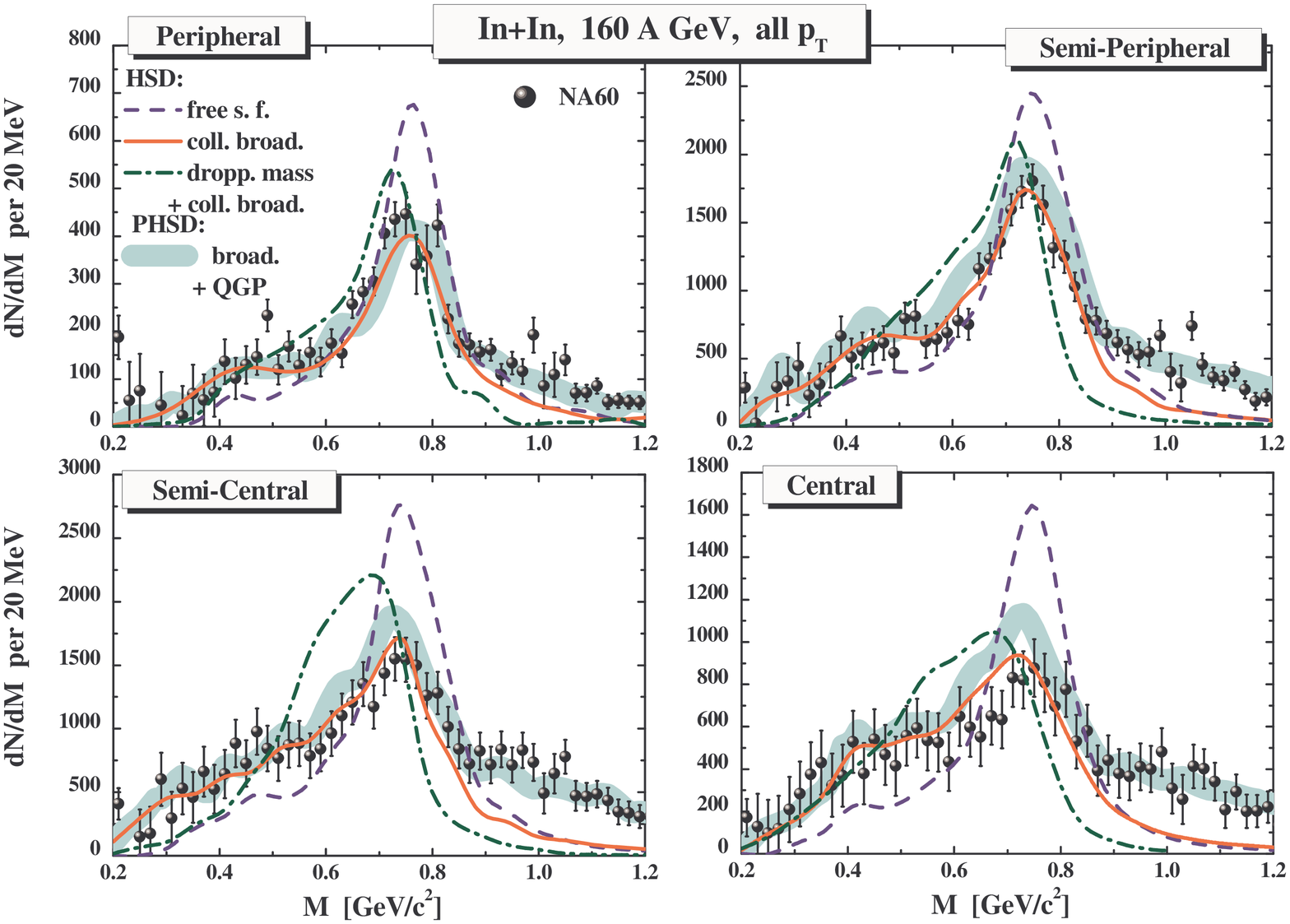}
    \label{ExcessSpectra}
    \caption{The HSD results for the mass differential dilepton
spectra from $In + In$ collisions at 158 A$\cdot$GeV in comparison
to the excess mass spectrum from NA60 \protect\cite{NA60}. The
actual NA60 acceptance filter and mass resolution have been
incorporated \cite{Sanja}.  The solid lines show the HSD results for
a scenario including the collisional bradening of the $\rho$-meson
whereas the dashed lines correspond to calculations with a 'free'
$\rho$ spectral functions for reference. The dash-dotted lines
represent the HSD calculations for the 'dropping mass + collisional
broadening' model. The bands represent the preliminary PHSD results
incorporating direct dilepton radiation from the QGP in addition to
a broadened $\rho$-meson.}
  \end{minipage}
  \hspace{0.02\textwidth}
  \begin{minipage}[b]{0.44\textwidth}
    \includegraphics[width=\textwidth]{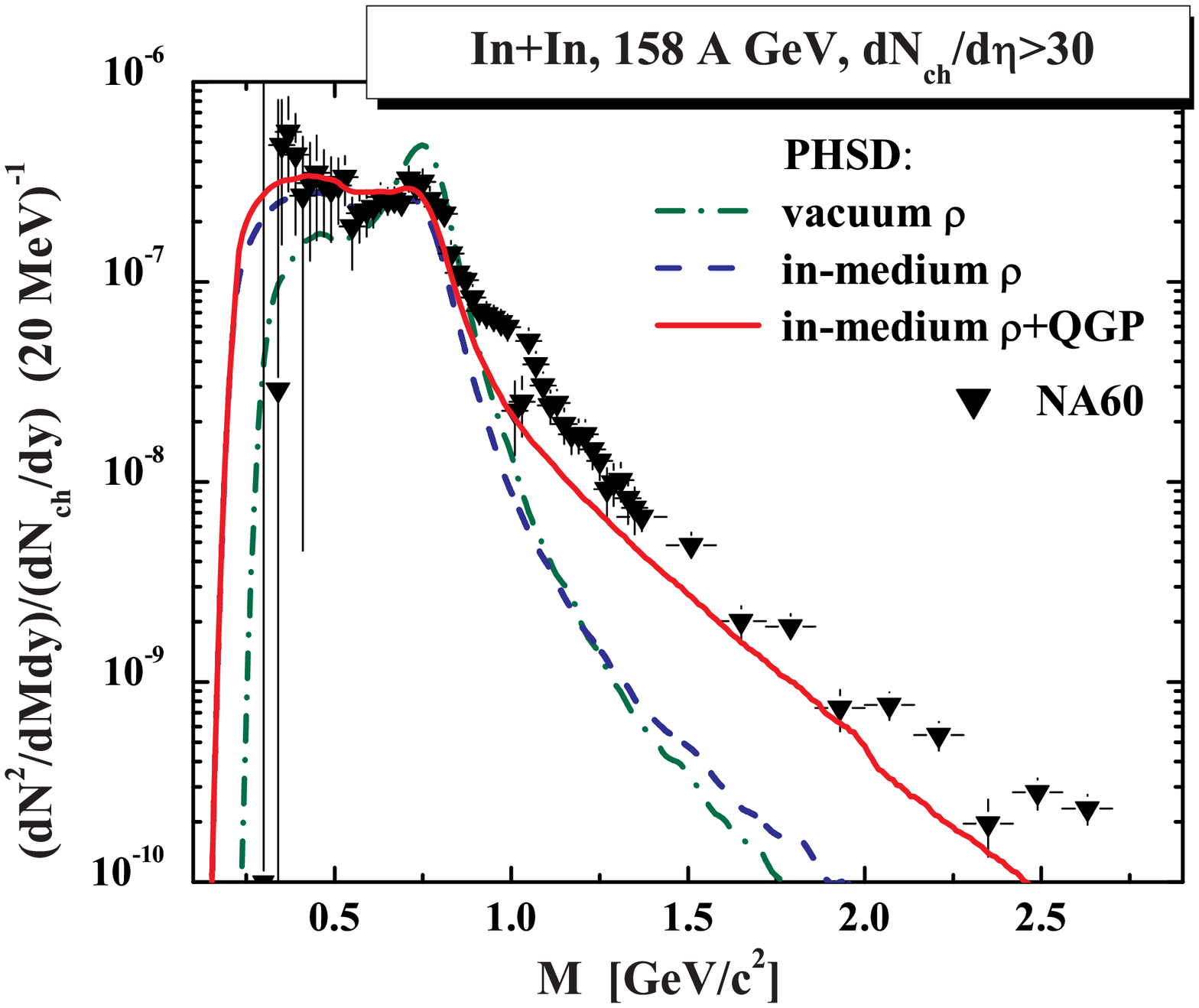}
    \label{NA60_AC}
    \caption{Acceptance corrected mass spectra of the excess dimuons
from $In+In$ at 158~AGeV integrated over $p_T$ in $0.2<p_T<2.4$~GeV from
PHSD compared to the data of NA60~\cite{Arnaldi:2008er}. The green
dash-dotted line shows the dilepton yield from the vacuum $\rho$
meson. The blue dashed line is the contribution to the dilepton
yield from the in-medium $\rho$ with broadened spectral function.
The red solid line presents the sum of the in-medium $\rho$ and QGP
dilepton radiation (the latter is calculated in the on-shell
approximation in this work).}
    \vspace{0.4cm}
  \end{minipage}
\end{figure*}

As we find in Fig. \ref{ExcessSpectra} the NA60 data favor the
scenario of the in-medium broadening of vector mesons.  Note that in
the data - presented in this plot - the D-meson contribution has not
been subtracted. The NA60 collaboration has published acceptance
corrected data with subtracted charm contribution
recently~\cite{Arnaldi:2008er}. In Fig.~\ref{NA60_AC} we present
PHSD results for the dilepton spectrum excess over the known
hadronic sources as produced in $In+In$ reactions at 158~AGeV
compared to the acceptance corrected data. We find that the spectrum
at invariant masses below 1~GeV is well reproduced by the $\rho$
meson yield, if a broadening of the meson spectral function in the
medium is assumed. On the other hand, the spectrum at $M>1$~GeV is
shown to be dominated by the partonic sources.

\begin{figure*}
  \begin{minipage}[b]{0.54\textwidth}
    \includegraphics[width=\textwidth]{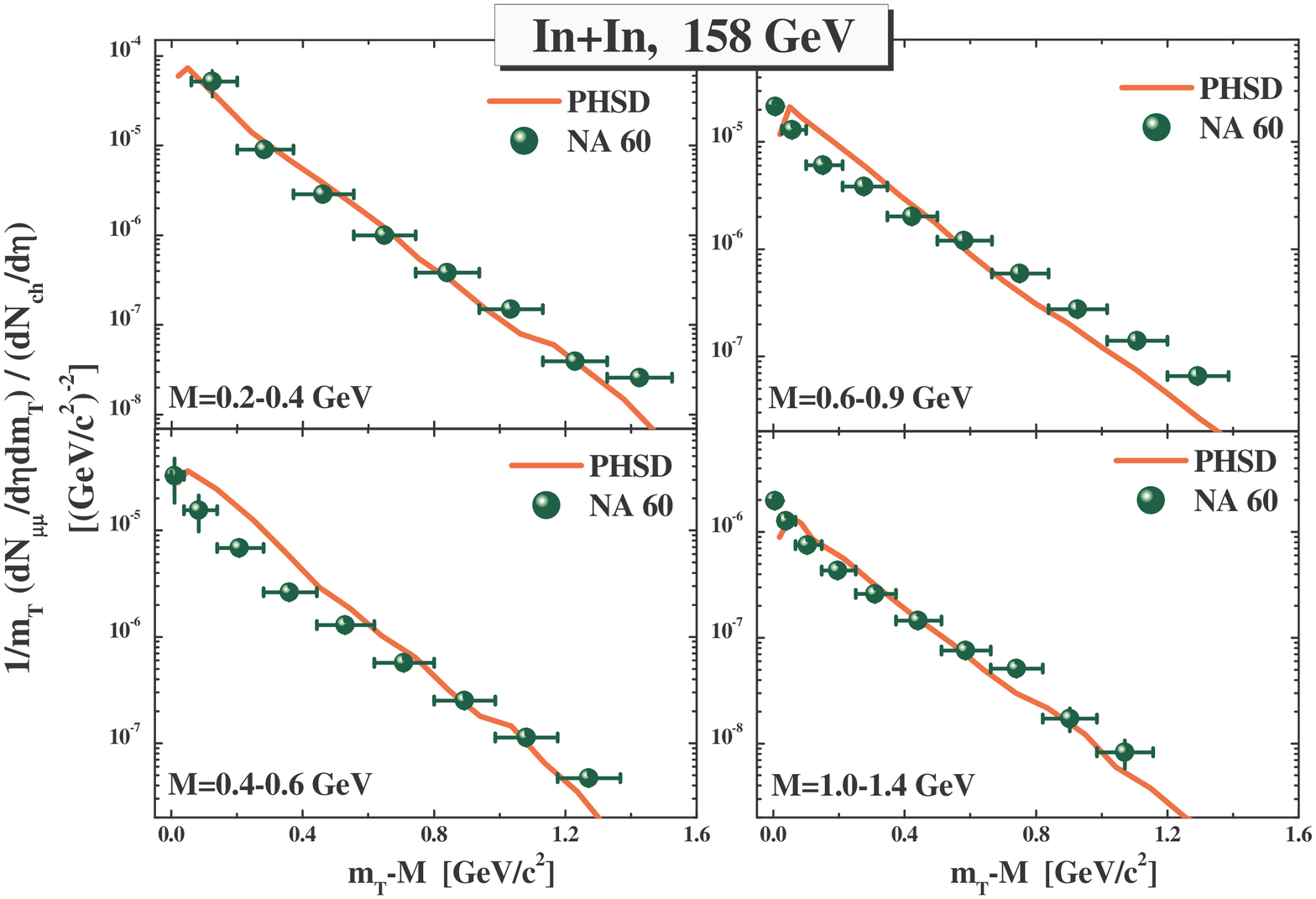}
  \end{minipage}
  \begin{minipage}[b]{0.44\textwidth}
    \includegraphics[width=\textwidth]{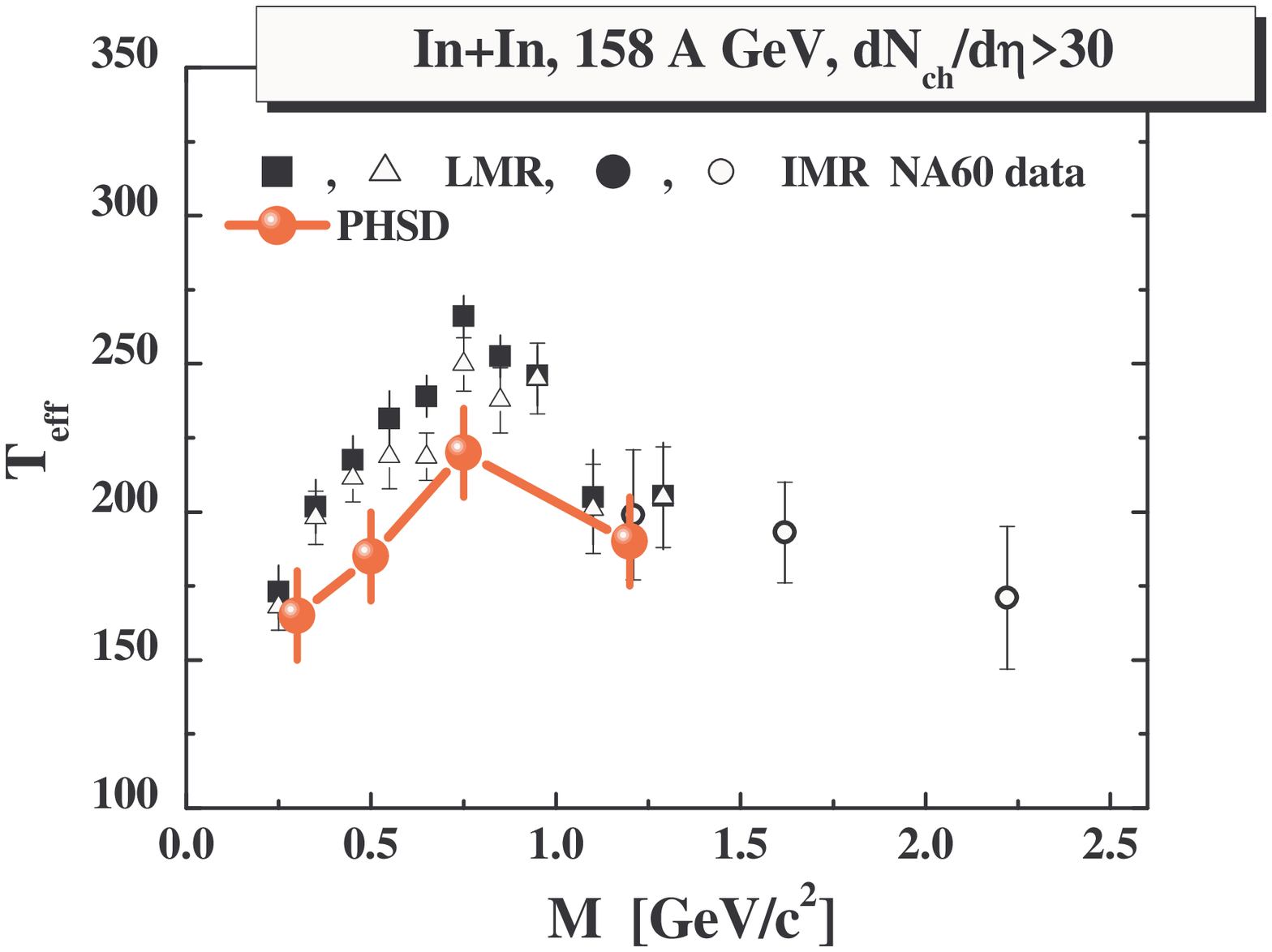}
  \end{minipage}
   \caption{{\bf Left:} Transverse mass spectra of dileptons for In+In at
158 A$\cdot$GeV. {\bf Right:} {\bf Right:} The inverse slope
parameter $T_{eff}$ of the dimuon yield from In+In at 158
A$\cdot$GeV as a function of the dimuon invariant mass in PHSD
compared to the data of the NA60
Collaboration~\protect{\cite{NA60,Arnaldi:2008er}}.} \label{Slopes}
\end{figure*}

It is also interesting to note that accounting for partonic dilepton
sources allows to reproduce in PHSD (cf. Fig.~\ref{NA60_AC}, rhs)
the intriguing finding of the NA60
Collaboration~\cite{NA60,Arnaldi:2008er} that the effective
temperature of the dileptons (slope parameters) in the intermediate
mass range is lower than the $T_{eff}$ of the dileptons from the
hadronic phase. The softening of the transverse mass spectrum with
growing invariant mass reflects that the partonic channels occur
dominantly before the collective radial flow has developed. This
feature of the data is also reproduced in PHSD (cf.
Fig.~\ref{Slopes}, rhs). A detailed look at the PHSD results shows
that in total we still underestimate the slope parameter $T_{eff}$
which might be due to missing partonic initial state effects or an
underestimation of partonic flow in the initial phase of the
reaction.

\section{RHIC energies}

\begin{figure*}
  \begin{minipage}[b]{0.49\textwidth}
    \includegraphics[width=\textwidth]{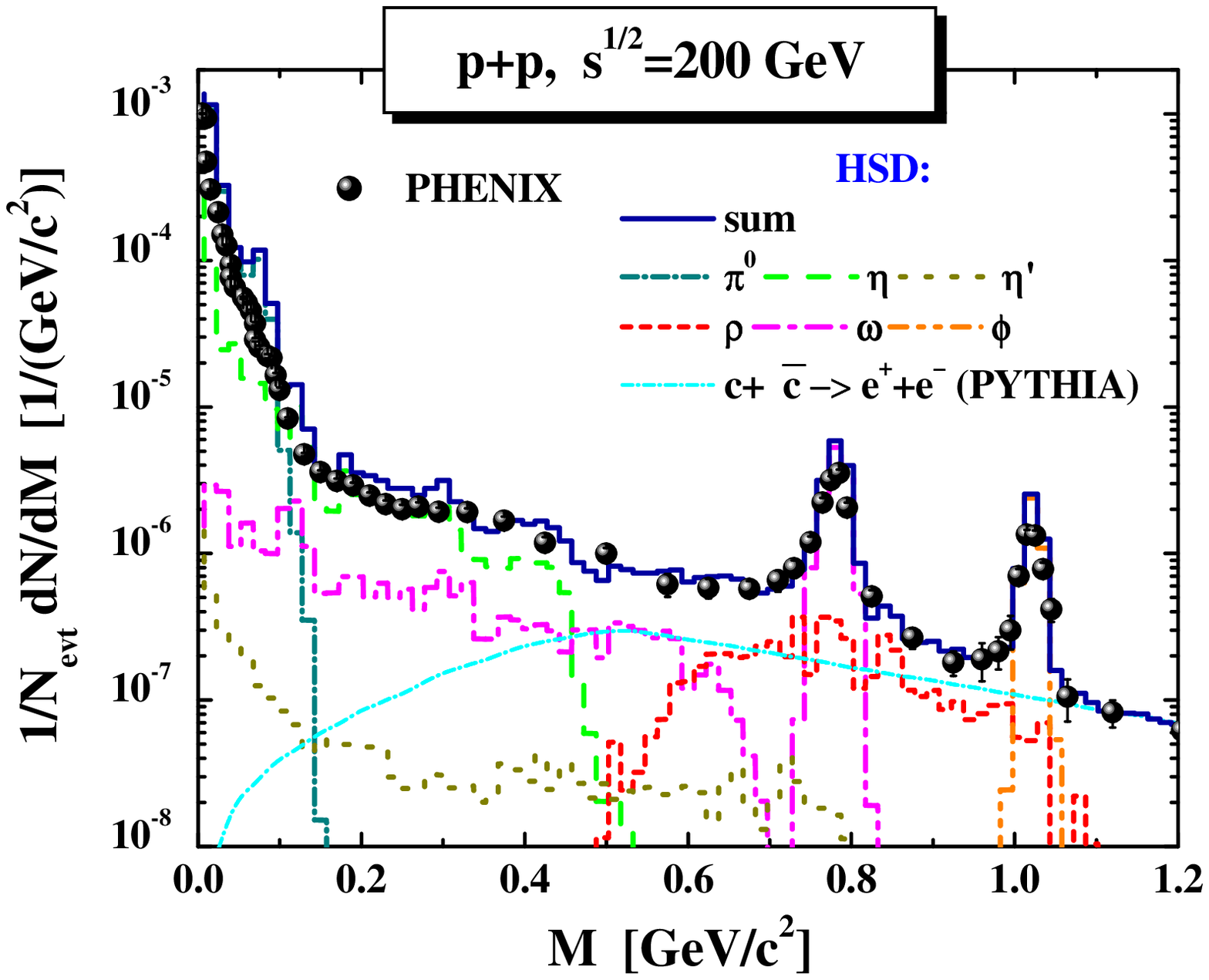}
    \caption{The HSD results  for the mass differential dilepton spectra
in case of $pp$ collisions at $\sqrt{s}$ = 200 GeV  in comparison to
the data from PHENIX~\protect{\cite{PHENIXpp}}. The actual PHENIX
acceptance and mass resolution have been incorporated (see legend
for the different color coding of the individual channels). Figure
taken from~\protect{\cite{here}}.} \label{pp}
     \vspace{2.3cm}
  \end{minipage}
  \hspace{0.02\textwidth}
  \begin{minipage}[b]{0.49\textwidth}
    \includegraphics[width=\textwidth]{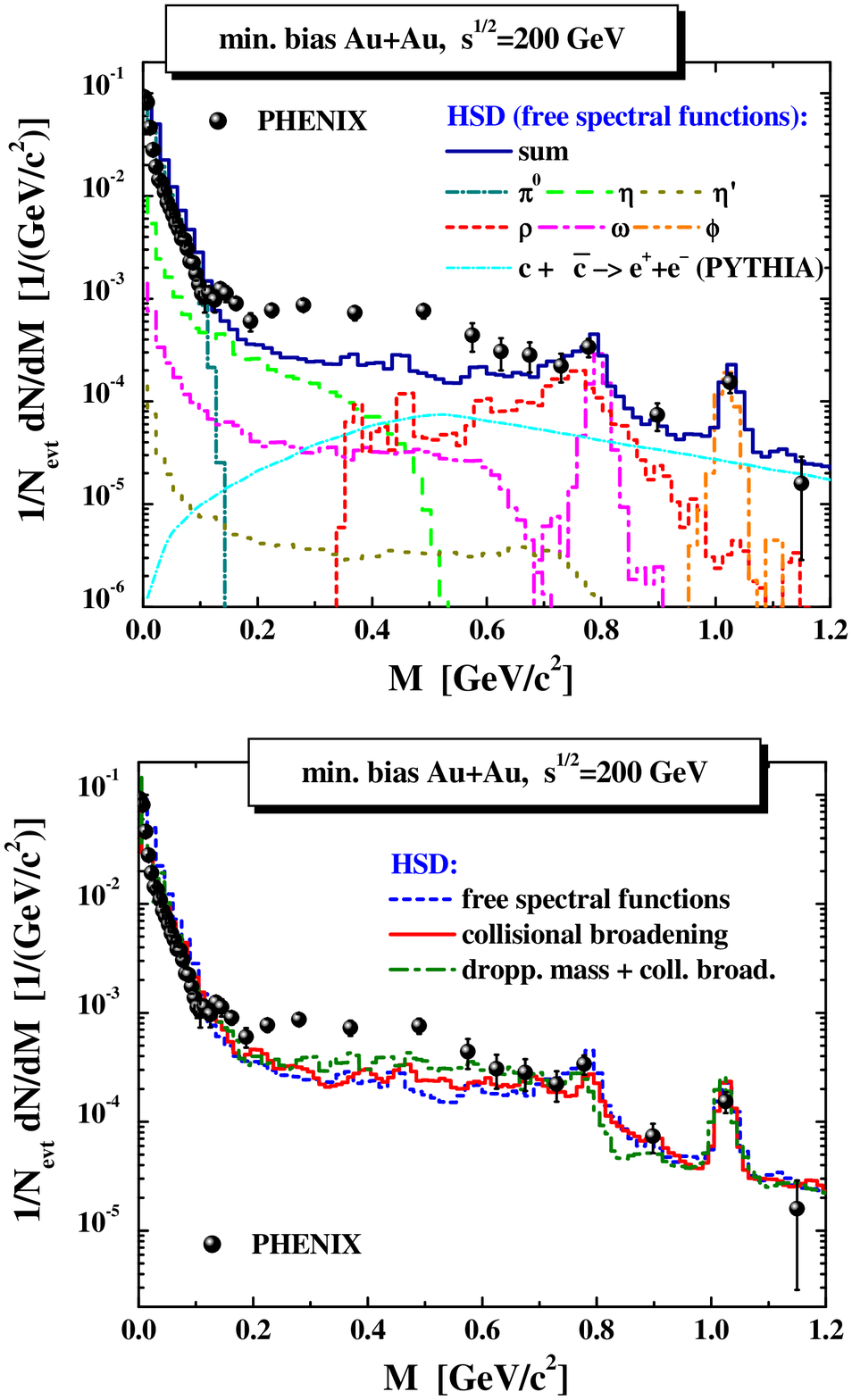}
  \end{minipage}
   \caption{The HSD
results  for the mass differential dilepton spectra in case of
inclusive $Au + Au$ collisions at $\sqrt{s}$ = 200 GeV in comparison
to the data from PHENIX~\protect{\cite{PHENIX}}. The actual PHENIX
acceptance filter and mass resolution have been
incorporated~\protect{\cite{Alberica}}. In the upper part the
results are shown for  vacuum spectral functions (for $\rho, \omega,
\phi$) including the channel decompositions (see legend for the
different color coding of the individual channels). The lower part
shows a comparison for the total $e^+e^-$ mass spectrum in case of
the 'free' scenario (dashed line), the 'collisional broadening'
picture (solid line) as well as the 'dropping mass + collisional
broadening' model (dash-dotted line). Figure taken
from~\protect{\cite{here}}. } \label{Fig3}
\end{figure*}

In 2008, the PHENIX Collaboration has presented first dilepton data
from $pp$ and $Au+Au$ collisions at Relativistic-Heavy-Ion-Collider
(RHIC) energies of $\sqrt{s}$ = 200 GeV \cite{PHENIX} which show an
even larger enhancement in $Au+Au$ reactions (relative to $pp$
collisions) in the invariant mass regime from 0.15 to 0.6 GeV than
the data at SPS energies \cite{NA60,CERES2}. We recall that HSD
provides a reasonable description of hadron production in $Au+Au$
collisions at $\sqrt{s}$ = 200 GeV~\cite{Brat03} such that we can
directly continue with the results for $e^+e^-$ pairs which are
shown in Fig.~\ref{pp} for $p+p$ collisions. We find that the
dilepton radiation in the elementary channel is well under control
in HSD, which is equivalent to PHSD for $p+p$ reactions.

We step on with the case of inclusive $Au + Au$ collisions in
comparison to the data from PHENIX~\cite{PHENIX} (Fig~\ref{Fig3}).
When including the in-medium modification scenarios for the vector
mesons, we achieve a sum spectrum which is only slightly enhanced
compared to the 'free' scenario. Whereas the total yield  is quite
well described in the region of the pion Dalitz decay as well as
around the $\omega$ and $\phi$ mass, HSD clearly underestimates the
measured spectra in the regime from 0.2 to 0.6 GeV by an average
factor of $\sim 3$. The low mass dilepton spectra from $Au+Au$ collisions
at RHIC (from the PHENIX Collaboration) are clearly underestimated
in the invariant mass range from 0.2 to 0.6 GeV in the 'collisional
broadening' scenario as well as in the 'dropping mass + collisional
broadening' model, i.e. when assuming a shift of the vector meson
mass poles with the baryon density. We mention that our results for
the low mass dileptons are very close to the calculated spectra from
van Hees and Rapp \cite{rapp3} as well as Dusling and 
Zahed~\cite{Dussi} (cf. the comparison in Ref.~\cite{AToia}).

Consequently we attribute this additional low mass enhancement seen
by PHENIX to non-hadronic sources. This assumption has to be tested
quantitatively. However, in order to make quantitative predictions
at experimentally relevant low dilepton mass ($\le 0.6$~GeV)
and strong coupling ($\alpha_S \! \sim \! 0.5 \! \div \! 1$), we
have to take into account not only the higher order pQCD reaction
mechanisms, but also the non-perturbative spectral functions
and self-energies of the quarks, anti-quarks and gluons thus going
beyond the leading twist.

For this purpose, off-shell cross sections were derived
in~\cite{olena2010} for dilepton production in the reactions $q+\bar
q\to l^+l^-$ (Drell-Yan mechanism), $q+ \bar q\to g+l^+l^-$ (quark
annihilation with the gluon Bremsstrahlung in the final state),
$q(\bar q)+g\to q(\bar q)+ l^+l^-$ (gluon Compton scattering), $g\to
q+\bar q+l^+l^-$ and $q(\bar q)\to q(\bar q)+g+l^+l^-$ (virtual
gluon decay, virtual quark decay) in the sQGP in effective
perturbation theory by dressing the quark and gluon lines in the
diagrams in Fig.~1 with the DQPM propagators for quarks and gluons.
The DQPM describes QCD properties in terms of single-particle
Green's functions (in the sense of a two-particle irreducible
approach) and leads to the notion of the constituents of the sQGP
being effective quasiparticles, which are massive and have broad
spectral functions (due to large interaction rates).

\begin{figure*}
  \begin{minipage}[b]{0.49\textwidth}
    \includegraphics[width=.95\textwidth]{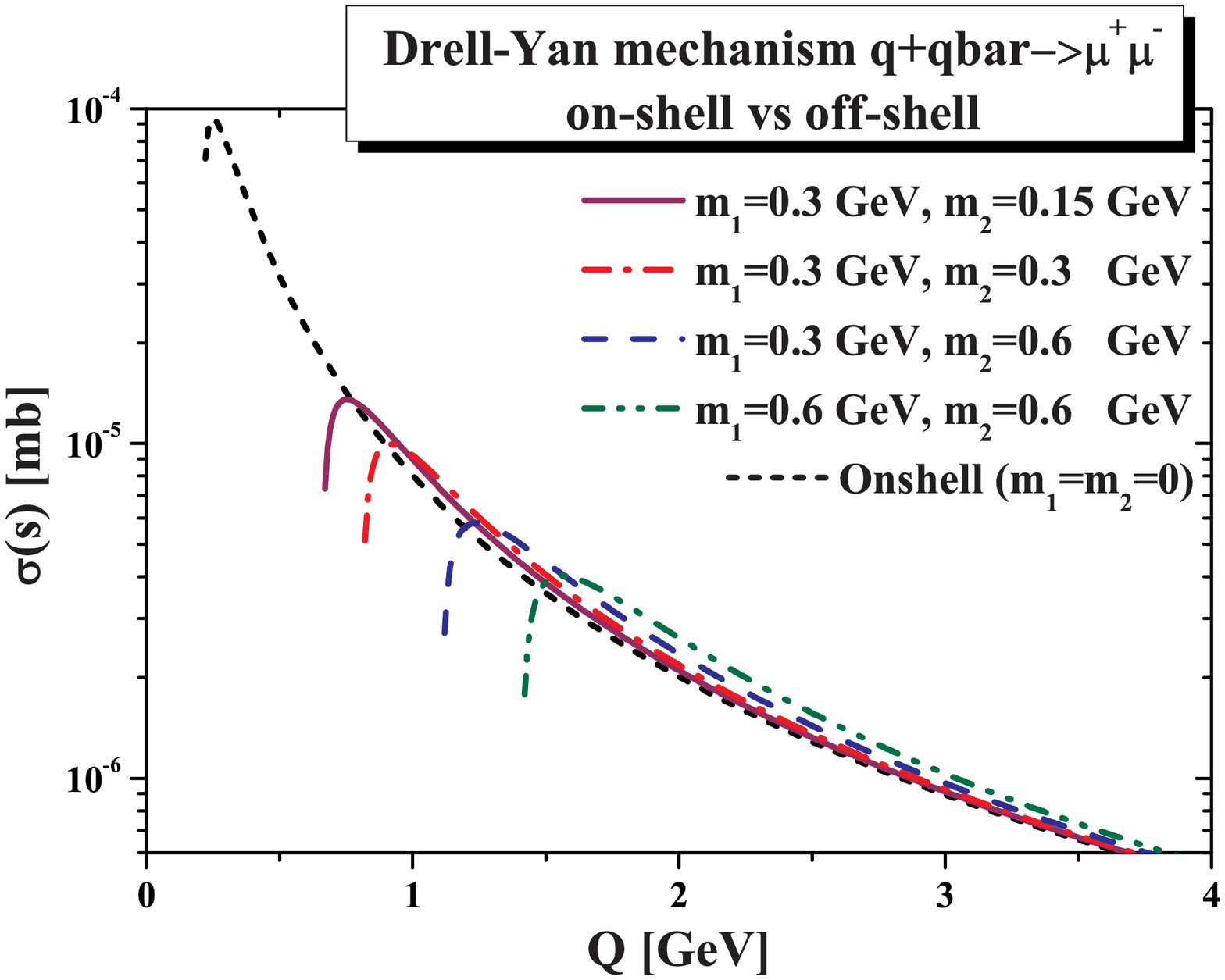}
    \caption{Dilepton production cross sections in the
Drell-Yan channel ($q+\bar q\to \mu^+ + \mu^-$). The short dashes
(black) line shows the on-shell, i.e. the standard pQCD result. The
other lines show the off-shell cross section, in which the
annihilating quark and antiquark have finite masses $m_1$ and $m_2$
with different values: $m_1=0.3$~GeV, $m_2=0.15$~GeV (solid magenta
line), $m_1=0.3$~GeV, $m_2=0.3$~GeV (dash-dotted red line),
$m_1=0.3$~GeV, $m_2=0.6$~GeV (dashed blue line), $m_1=0.6$~GeV,
$m_2=0.6$~GeV (dash-dot-dot green line).}
    \label{DYoffVsOn}
  \end{minipage}
  \hspace{0.02\textwidth}
  \begin{minipage}[b]{0.49\textwidth}
    \includegraphics[width=1\textwidth]{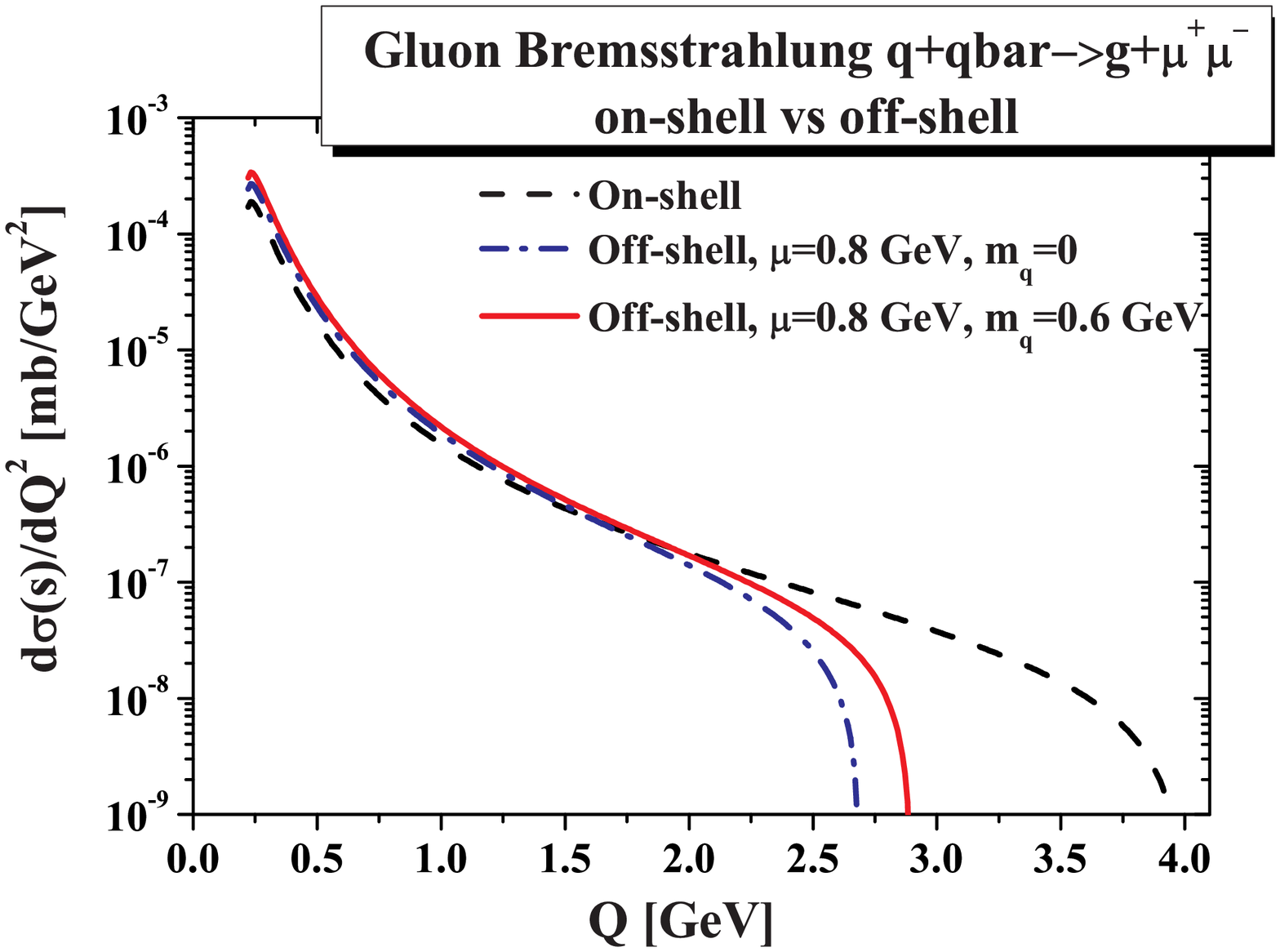}
    \caption{Comparison of off-shell and on-shell cross sections
     for dilepton production in the gluon Bremsstrahlung
     $q+\bar q\to g+\mu^+\mu^-$ channel at $\sqrt{s}=4$~GeV.
     The dashed black line shows the on-shell (pQCD)
     cross section regularized
     by the cut on the gluon mass $\mu_{cut}=0.206$~GeV,
     the blue dash-dotted line presents the off-shell cross section for
     the gluon mass fixed to $\mu=0.8$~GeV and on-shell quark and
     anti-quark ($m_1=m_2=m_3=0$). The red solid line gives the off-shell
     result for $\mu=0.8$~GeV, $m_1=m_2=m_3=m_q=0.6$~GeV.  }
    \label{gBRoffshell}
  \end{minipage}
 \end{figure*}

Dilepton production cross sections in the Drell-Yan mechanism are
plotted on the l.h.s. of Fig.~\ref{DYoffVsOn}. The short dashes
(black) line shows the on-shell, i.e. the standard perturbative
result. The other lines show the off-shell cross section, in which
the annihilating quark and antiquark have finite masses $m_1$ and
$m_2$ with different values: $m_1=0.3$~GeV, $m_2=0.15$~GeV (solid
magenta line), $m_1=0.3$~GeV, $m_2=0.3$~GeV (dash-dotted red line),
$m_1=0.3$~GeV, $m_2=0.6$~GeV (dashed blue line), $m_1=0.6$~GeV,
$m_2=0.6$~GeV (dash-dot-dot green line).

In Fig.~\ref{gBRoffshell}, the off-shell cross sections for the
quark annihilation with gluon bremsstrahlung channel at various
values of quark and gluon off-shellnesses (masses) are compared to
the on-shell (pQCD) result. The dashed black line shows
  the on-shell cross section regularized by a cut on gluon mass
  $\mu_{cut}=0.206$~GeV, the red solid line
  presents the off-shell cross section for the gluon mass fixed to
  $\mu=0.8$~GeV and on-shell quark and anti-quark ($m_1=m_2=m_3=0$).
  The blue dash-dotted line gives the off-shell result for $\mu=0.8$~GeV,
  $m_1=m_2=m_3=m_q=0.6$~GeV.
One readily notices the shift of the maximum allowed mass of the
lepton pair to a lower value (in order to produce a massive gluon in
the final state).

By implementing the off-shell partonic processes into the PHSD
transport approach, a consistent calculation of the dilepton
production in heavy-ion collisions at RHIC energies will be performed
in near future. The comparison to the dilepton data double
differentially in mass and $p_T$ will open the possibility to study
the relative importance of different processes in the dilepton
production and guide us towards a better understanding of the
properties of matter at high densities and temperatures
as created in heavy-ion collisions.

\section{Summary}

Dilepton production in relativistic heavy-ion collisions is a
valuable probe of two phenomena of fundamental interest: the vector
meson properties in a hot and dense medium and of the quark and
gluon dynamics in the deonfined state of matter (sQGP). The Parton
Hadron String Dynamics~\cite{CasBrat} (PHSD) transport approach
incorporates the relevant off-shell dynamics of the vector mesons as
well as the explicit partonic phase in the early hot and dense
reaction region.

A comparison of the transport calculations to the data of the CERES
and NA60 Collaborations points towards 'melting' modification of the
$\rho$'-meson at high densities, i.e. the broadening of the vector
meson's spectral function. On the other hand, the spectrum for
$M>1$~GeV is shown to be dominated by the partonic sources.

The low mass dilepton spectra from $Au+Au$ collisions at RHIC (from
the PHENIX Collaboration) are clearly underestimated by the hadronic
channels in the invariant mass range from 0.2 to 0.6 GeV in the
'collisional broadening' scenario as well as in the 'dropping mass +
collisional broadening' model, i.e. when assuming a shift of the
vector meson mass poles with the baryon density. We attribute this
additional low mass enhancement seen by PHENIX to radiation from
the strongly interacting QGP.
This assumption has to be tested quantitatively. Strong interaction
of partons - reflected in their high masses and broad widths - lead
to higher-twist corrections to the perturbative cross sections
especially at the experimentally relevant low dilepton masses ($\le
0.6$~GeV). By implementing the off-shell partonic processes into the
PHSD transport approach, a consistent calculation of the dilepton
production in heavy-ion collisions at RHIC energies will be performed
in near future that will allow us answer whether the observed access can be
attributed to the radiation by off-shell partons.


\section*{Acknowledgments}

OL acknowledges financial support within the ``HIC for FAIR" center
of the ``LOEWE'' program.


\section*{References}
\begin{thebibliography}{9}
\bibitem{CERES}
        G. Agakichiev {\it et al.}, CERES Collaboration,
              \emph{ Phys. Rev. Lett.} {\bf 75} (1995) 1272;
        Th. Ullrich {\it et al.},  \emph{ Nucl. Phys.} { A}{\bf 610} (1996) 317c;
        A. Drees, \emph{ Nucl. Phys.} { A}{\bf 610} (1996) 536c.
\bibitem{HELIOS}
        M. A. Mazzoni, HELIOS Collaboration,
              \emph{ Nucl. Phys.} { A}{\bf 566} (1994) 95c;
        M. Masera, \emph{ Nucl. Phys.} { A}{\bf 590} (1995) 93c;
        T. {\AA}kesson et al., \emph{ Z. Phys.} { C}{\bf 68} (1995) 47.
\bibitem{Linnyk:2009nx}
  O.~Linnyk, E.~L.~Bratkovskaya and W.~Cassing,
  \emph{Nucl.\ Phys.}\  A {\bf 830} (2009) 491C.
\bibitem{CasBrat} W. Cassing and E. L. Bratkovskaya,
    \emph{Phys. Rev.} C {\bf 78} (2008) 034919;
  \emph{Nucl.\ Phys.}\  A {\bf 831} (2009) 215.
\bibitem{CBRep98}
        W. Cassing, E. L. Bratkovskaya,
        \emph{ Phys. Rept.} {\bf 308} (1999) 65.
\bibitem{Brat97}
        E. L. Bratkovskaya, W. Cassing,
      \emph{ Nucl. Phys.} { A}{\bf 619} (1997) 413.
\bibitem{Ehehalt}
    W. Ehehalt, W. Cassing, \emph{ Nucl. Phys.} { A }{\bf 602} (1996) 449.
\bibitem{Cass_off1}
    W. Cassing, S. Juchem, \emph{ Nucl. Phys.} { A }{\bf 665} (2000)
    377; {\it ibid.} { A }{\bf 672} (2000) 417.
\bibitem{MuellerSumRules}
  J.~Ruppert, T.~Renk and B.~M{\"u}ller,
  \emph{Phys.\ Rev.}\  C {73} (2006)  034907.
\bibitem{Brat08}
       E. L. Bratkovskaya, W. Cassing,
      \emph{Nucl. Phys.} A {\bf 807} (2008) 214.
\bibitem{here}
  E. L. Bratkovskaya, W. Cassing and O. Linnyk,
  \emph{Phys.\ Lett.}\  B {\bf 670} (2009) 428.
\bibitem{Cassing06}
      W. Cassing, \emph{Nucl. Phys.} A 791 (2007) 365; {\it ibid.}  A 795 (2007) 70.
\bibitem{NA60}
      R. Arnaldi {\it et al.}, NA60 Collaboration,
              \emph{Phys. Rev. Lett.} {\bf 96} (2006)   162302;
      J. Seixas {\it et al.},  \emph{J. Phys.} G {\bf 34} (2007) S1023;
       S. Damjanovic {\it et al.}, \emph{Nucl. Phys.} {\bf A 783} (2007) 327c.
\bibitem{CERES2}
       D. Adamova {\it et al.} CERES Collaboration,
         \emph{Nucl. Phys.} A {\bf 715} (2003) 262;
         \emph{Phys. Rev. Lett.} {\bf 91} (2003) 042301;
       G. Agakichiev {\it et al.},
         \emph{Eur. Phys. J.} C {\bf 41} (2005) 475;
       D. Adamova {\it et al.}
  \emph{Phys.\ Lett.}\  B {\bf 666} (2008) 425;
       A. Marin {\it et al.}; Proceedings of CPOD07,
              PoS 034 (2007).
\bibitem{Arnaldi:2008er} 
  R.~Arnaldi {\it et al.}, NA60 Collaboration,
  Eur.\ Phys.\ J.\  C {\bf 59}, 607 (2009)
\bibitem{PHENIXpp}
       A. Adare {\it et al.}, PHENIX Collaboration,
  \emph{Phys.\ Lett.}\  B {\bf 670} (2009) 313
\bibitem{PHENIX}
       A. Toia {\it et al.},  PHENIX Collaboration,
       \emph{Nucl. Phys.} A {\bf 774} (2006) 743;
       \emph{Eur. Phys.} J {\bf 49} (2007) 243;
      S. Afanasiev {\it et al.}, PHENIX Collaboration,  arXiv:0706.3034 [nucl-ex];
   A.~Adare {\it et al.}, PHENIX Collaboration,
  arXiv:0912.0244 [nucl-ex].
\bibitem{Alberica} A. Toia, private communication.
\bibitem{Brat03}
       E. L. Bratkovskaya, W. Cassing, H. St\"ocker,
       \emph{Phys. Rev. } {\bf C 67}  (2003) 054905;
       E. L. Bratkovskaya {\it et al.}
       \emph{Phys. Rev. } {\bf C 69} (2004) 054907.
\bibitem{rapp3}
       H. van Hees, R. Rapp, Phys. Rev. Lett. {\bf 97} (2006) 102301.
\bibitem{Dussi}
  K.~Dusling and I.~Zahed,
  \emph{Nucl.\ Phys.}\  A {\bf 825} (2009) 212.
\bibitem{AToia}
  A.~Toia,
  \emph{J.\ Phys.}\ G {\bf 35} (2008) 104037.
\bibitem{Sanja}
      S.~Damjanovic, private communication.
\bibitem{olena2010}
      O.~Linnyk, arXiv:1004.2591 [hep-ph].
\end{thebibliography}
\end{document}